\documentclass[superscriptaddress,twocolumn,showpacs,aps,nofootinbib,notitlepage,preprintnumbers]{revtex4-1}

\pdfoutput=1

\usepackage{aas_macros}
\usepackage[utf8]{inputenc}
\usepackage{natbib}
\usepackage{amsmath}
\usepackage{amssymb}
\usepackage{amsfonts}
\usepackage{bm}  % for more advanced \bm
\usepackage{xfrac}
\usepackage{graphicx}
\usepackage[justification=RaggedRight]{caption}
\usepackage{subcaption}
\usepackage{hyperref}
\usepackage{listings}
\lstdefinelanguage{yaml}{
  sensitive=true,
  morecomment=[l]{\#}}
\usepackage[dvipsnames]{xcolor}
\usepackage{todonotes}
\usepackage[normalem]{ulem}
\usepackage{pythonhighlight}
\colorlet{keywordcolour}{Maroon!90!black}
\colorlet{stringcolour}{Periwinkle!60!black}
\colorlet{literatecolour}{Mahogany!90!black}

\newcommand{\defeq}{:=}

\newcommand{\eqpunctsp}{\;}
\newcommand{\eqc}{\eqpunctsp ,}
\newcommand{\eqp}{\eqpunctsp .}

\renewcommand{\vector}[1]{\bm {#1}}

\newcommand{\probn}{\mathcal{P}}
\newcommand{\prob}[1]{\probn\left(#1\right)}
\newcommand{\cond}{\,|\,}

\newcommand{\liken}{\mathcal{L}}
\newcommand{\like}[1]{\liken\left(#1\right)}
\newcommand{\priorn}{\pi}
\newcommand{\prior}[1]{\priorn\left(#1\right)}

\newcommand{\data}{\mathcal{D}}
\newcommand{\model}{\mathcal{M}}

\newcommand{\params}{\theta}

\newcommand{\normal}[2]{\mathcal{N}\left(#1,\,#2\right)}
\newcommand{\obs}{\vector{O}}
\usepackage{xspace}
\newcommand{\cobaya}{\texttt{Cobaya}\xspace}
\newcommand{\Python}{\texttt{Python}\xspace}
\newcommand{\cosmomc}{\texttt{CosmoMC}\xspace}
\newcommand{\montepython}{\texttt{MontePython}\xspace}
\newcommand{\cosmosis}{\texttt{CosmoSIS}\xspace}
\newcommand{\camb}{\texttt{CAMB}\xspace}
\newcommand{\class}{\texttt{CLASS}\xspace}

\newcommand{\polychord}{\texttt{PolyChord}\xspace}

\newcommand{\scipy}{\texttt{scipy}\xspace}
\newcommand{\pybobyqa}{\texttt{Py-BOBYQA}\xspace}
\newcommand{\bobyqa}{\texttt{BOBYQA}\xspace}
\newcommand{\pandas}{\texttt{Pandas}\xspace}
\newcommand{\getdist}{\texttt{GetDist}\xspace}

\newcommand{\numpy}{\texttt{numpy}\xspace}
\newcommand{\openblas}{\texttt{OpenBLAS}\xspace}

\begin{document}
%%%%%%%%%%%%%%%%%%%%%%%%%%%%%%%%%%%%%%%%%%%%%%%%%%%%%%%%%%%%%%%%%%%%%%%%%%%%%%%%%%%

\title{\cobaya: Code for Bayesian Analysis of hierarchical physical models}

\author{Jesús Torrado}
\affiliation{Department of Physics \& Astronomy, University of Sussex, Brighton BN1 9QH, UK}
\affiliation{Institute for Theoretical Particle Physics and Cosmology (TTK), RWTH Aachen University, D-52056 Aachen, Germany}
\author{Antony Lewis}
\affiliation{Department of Physics \& Astronomy, University of Sussex, Brighton BN1 9QH, UK}

\preprint{TTK-20-15}

\begin{abstract}
  We present \cobaya, a general-purpose Bayesian analysis code aimed at models with complex internal interdependencies.

  Without the need for specific code by the user, interdependencies between different stages of a model pipeline are exploited for sampling efficiency: intermediate results are automatically cached, and parameters are grouped in blocks according to their dependencies and optimally sorted, taking into account their individual computational costs, so as to minimize the cost of their variation during sampling, thanks to a novel algorithm.

  \cobaya allows exploration of posteriors using a range of Monte Carlo samplers, and also has functions for maximization and importance-reweighting of Monte Carlo samples with new priors and likelihoods.

  \cobaya is written in Python in a modular way that allows for extendability, use of calculations provided by external packages, and dynamical reparameterization without modifying its source.
    It can exploit hybrid OpenMP/MPI parallelization, and has sub-millisecond overhead per posterior evaluation.
    \\[.4em]
    Though \cobaya is a general purpose statistical framework, it includes interfaces to a set of cosmological Boltzmann codes and likelihoods (the latter being agnostic with respect to the choice of the former), and automatic installers for external dependencies.
\end{abstract}

\maketitle

%%%%%%%%%%%%%%%%%%%%%%%%%%%%%%%%%%%%%%%%%%%%%%%%%%%%%%%%%%%%%%%%%%%%%%%%%%%%%%%%%%%
\section{Introduction}
%%%%%%%%%%%%%%%%%%%%%%%%%%%%%%%%%%%%%%%%%%%%%%%%%%%%%%%%%%%%%%%%%%%%%%%%%%%%%%%%%%%

Bayesian parameter estimation is a problem usually encountered when testing physical models against some experimental data. In these cases, inference from the posterior of the model is often done using a sampling algorithm. The posterior's likelihood in realistic scenarios is usually a complex hierarchical model, involving costly computations of intermediate quantities and possibly multiple likelihoods from different data sets. Each part of the calculation may have its own parameter dependencies, from the underlying parameters of main physical interest to nuisance parameters describing the experimental model or uncertainties in different parts of the theoretical calculation. Different subsets of parameters may have very different computational costs, depending on which parts of the model have to be recalculated when each parameter is varied. If subsets of the parameters can be updated with low computational cost, so there is a significant speed hierarchy, this can be exploited to greatly improve sampling efficiency. However, the simplest applications of standard sampling algorithms usually fail to exploit this property.

\cobaya (COde for BAYesian Analysis; meaning ``guinea pig'' in Spanish) is a general framework for defining a pipeline of interdependent calculations, automatically exploiting the hierarchical nature of the model when sampling from its parameters. The main advantage of Cobaya resides in its ability to automatically analyse the dependency structure of the different model components, and to block parameters according to which parts of the computation (directly or indirectly) depend on them, sorting these parameters blocks in an optimal way so that looping over them takes the least amount of time and no part of the calculation is unnecessarily repeated (intermediate results are automatically cached). This also allows for oversampling of fast blocks according to their speed and use of sampling algorithms (such as fast dragging~\cite{Neal04}) that can efficiently exploit the presence of very fast parameter subspaces. \cobaya includes its own MCMC algorithms (adapted from \cosmomc \cite{Lewis:2013hha}) and an interface to the nested sampler \polychord \cite{2015MNRAS.450L..61H,2015MNRAS.453.4384H}. In addition to Monte Carlo samples, \cobaya includes a importance-weighting tool, and interfaces with popular minimizers for posterior and likelihood maximization (from \scipy \cite{scipy} and \pybobyqa implementation \cite{2018arXiv180400154C} of the \bobyqa algorithm \cite{Powell2009})\footnote{The structure is however general for samplers where posterior derivatives are not available: additional samplers could be used by providing a compatible sampler class implementation.} \cobaya is also ready for high-performance computation in clusters, featuring MPI parallelization and batch job generation.

Calculations can be separated into physically distinct but interdependent theoretical calculations and likelihood evaluations, where any part of the calculation can be provided by separately-maintained external Python packages each with its own parameter dependencies. This allows theory modellers and analysers of experimental data to provide and maintain different cleanly separated parts of the calculation rather than relying on a single monolithic package that does everything.

Though \cobaya is a completely generic framework, it includes out-of-the-box support for cosmological parameter estimation via a series of interfaces to a number of cosmological theory codes (\camb \cite{Lewis:1999bs,Howlett:2012mh} and \class \cite{Lesgourgues:2011re,Blas:2011rf})and data likelihoods. In that regard, it compares to \cosmomc \cite{Lewis:2002ah,Lewis:2013hha}, and later \montepython \cite{2013JCAP...02..001A,Brinckmann:2018cvx} and \cosmosis \cite{Zuntz:2014csq}, amongst others. \cobaya does not re-distribute most of the original likelihoods, data sets or theory codes, but provide installers for stock distributions. \cobaya can also be used as a simple cosmological likelihood wrapper, providing stand-alone cosmological posteriors that can be incorporated into the user's own statistical or machine-learning pipeline.

\cobaya is distributed as a standard \Python package\footnote{\url{https://pypi.org/project/cobaya/}}, which allows for easy management of dependencies and simple roll-out of updates. Extensive documentation is provided at \url{https://cobaya.readthedocs.io}. Users are also welcome to contribute through the central git repository at \url{https://github.com/CobayaSampler/cobaya}.

In the rest of the paper, we describe the probabilistic model underlying \cobaya's framework (Sec.~\ref{sec:probmodel}), comment on our main goals and consequent design choices (Sec.~\ref{sec:goals}), and describe the main structure of the code (Sec.~\ref{sec:structure}). We then briefly discuss use on high-performance computers (Sec.~\ref{sec:hpc}) and describe the specific realistic case of using \cobaya\ for cosmological data analysis (Sec.~\ref{sec:cosmo}). We explain the technical details of our new optimized parameter blocking in Appendix~\ref{app:blocking}, and display some of the source code necessary to reproduce the used case presented in Sec.\ \ref{sec:cosmo} in Appendix \ref{sec:cosmo_demo_source}.

%%%%%%%%%%%%%%%%%%%%%%%%%%%%%%%%%%%%%%%%%%%%%%%%%%%%%%%%%%%%%%%%%%%%%%%%%%%%%%%%%%%
\section{Probabilistic model}\label{sec:probmodel}
%%%%%%%%%%%%%%%%%%%%%%%%%%%%%%%%%%%%%%%%%%%%%%%%%%%%%%%%%%%%%%%%%%%%%%%%%%%%%%%%%%%

Bayesian inference on model parameters is usually done by performing Monte Carlo sampling on the posterior $\prob{\params\cond\data,\model}$, where $\model$ is some model parameterized by $\theta$ following some \emph{prior} $\prior{\theta\cond\model}$, and $\data$ is some data set we aim to model using $\model$. The probability for observing the data $\data$ given some particular set of model parameters is the \emph{likelihood}  $\like{\data\cond\model(\theta)}$. These quantities are related through the \emph{Bayes theorem}:
\begin{equation}
  \label{eq:bayestheorem}
  \prob{\params\cond\data,\model} =
  \frac{\like{\data\cond\model(\theta)}\prior{\theta\cond\model}}
       {\prob{\data\cond\model}}
  \eqc
\end{equation}
where $\prob{\data\cond\model}$ is the marginal likelihood or \emph{evidence}, which equals the integral of the numerator in the right-hand side.

In simple physical applications, the data that appears in the likelihood is expressed in terms of some summary \emph{observable} $\obs_\data$. In those cases, the likelihood consists of a comparison between the observed $\obs_\data$ and the one computed from a deterministic \emph{theoretical model} $\mathcal{T}$, $\obs_\mathcal{T}$, filtered through the simulated pipeline $\mathcal{F}_\mathcal{E}$ of the experiment following some \emph{experimental model} $\mathcal{E}$. For a simple Gaussian likelihood, this would look like
\begin{equation}
  \label{eq:like}
  \obs_\data \sim \normal{\mathcal{F}_\mathcal{E}(\obs_\mathcal{T})}{\Sigma_{\mathcal{E},\mathcal{T}}}
  \eqc
\end{equation}
where the covariance matrix $\Sigma_{\mathcal{E},\mathcal{T}}$, dependent on both the theoretical and the experimental model, parameterizes those deviations.

This simple example illustrates how a natural \emph{speed hierarchy} appears in the parameter space: whenever we vary parameters of the theoretical model, we would have to recompute both the theoretical prediction for the observable $\obs_\mathcal{T}$ \emph{and} the likelihood, whereas a variation in the experimental model parameters only requires recomputation of the likelihood (as long as we cache and reuse the computed $\obs_\mathcal{T}$). The likelihood calculation for fixed theoretical model is also often much faster than computing the theoretical prediction $\obs_\mathcal{T}$. Exploiting this speed hierarchy efficiently can save a considerable amount of time when sampling from the posterior (see appendix \ref{app:blocking}).

In general, physical inference pipelines may have several components and dependencies, without a sharp distinction between a theoretical and an experimental model. In \cobaya we allow the experimental likelihoods to communicate with one or more theory codes (and possibly other likelihoods) to calculate the required observables and any necessary derived parameters. Theory codes can depend on sampled parameters and on the outputs of other theory codes, so we allow for general (non-circular) dependencies between the different components so that the calculation can be cleanly modularized. This also allows any resulting hierarchy in the speed of the likelihood calculation under changes in different parameter variations to be exploited efficiently.

When various theoretical codes are available that can produce the same quantities, interfaces can be choice-agnostic so that they can be exchanged easily for testing and comparison.

%%%%%%%%%%%%%%%%%%%%%%%%%%%%%%%%%%%%%%%%%%%%%%%%%%%%%%%%%%%%%%%%%%%%%%%%%%%%%%%%%%%
\section{Goals and design choices}\label{sec:goals}
%%%%%%%%%%%%%%%%%%%%%%%%%%%%%%%%%%%%%%%%%%%%%%%%%%%%%%%%%%%%%%%%%%%%%%%%%%%%%%%%%%%

We aimed at the following development and usability goals, and made design choices accordingly:

\paragraph{Modularity:} The different actors in the Bayesian model (prior, likelihood, theory model, Monte Carlo sampler), are individual objects which share as little information as possible: prior and sampler know which parameters are sampled, but they do not know about any other fixed parameter of the model; the likelihood, in turn, deals only with input/output parameters and observables, not caring about their prior; the sampler does now know which individual likelihood (or theory code) understands which parameters, and lets a \emph{model} wrapper manage that. These design choices impose some compromises, but are a fair price for making the code more easily extendable and maintainable.

\paragraph{Rapid prototyping:} We have attempted to lower as much as possible the barrier to adding new priors, likelihoods, theory codes or samplers, or modify existing ones, all without touching \cobaya's source code or having to write a lot of \textit{wrapping} code. In order to achieve this, we have developed an API (i.e.\ \emph{application programming interface}) that offloads the need by the user to write specific code to handle communication between different parts of the pipeline, to cache intermediate results, or to exploit the parameter speed hierarchy. The user just needs to describe what a pipeline component needs and provides, and the actual calculation of the provided quantities. Creating such an API has been made much easier by writing our code in \Python, which allows us to discover parameter dependencies of generic functions (used as priors, likelihoods or parameter redefinitions) via its \textit{introspection} capabilities. Python is able to interface with libraries written in other languages, allowing use of external compiled theory codes and likelihoods, and also has useful computational capabilities (e.g.\ handling infinities directly). Though Python, as in interpreted language, is in general slower than compiled languages such as C or Fortran, this will not necessarily have a significant impact on computational speed, since most heavy computations are usually performed by calling functions in optimized packages based on a compiled back-end library (such as \numpy). Not-easy-to-offload computations (e.g.\ those containing long explicit loops) can also be implemented using popular just-in-time compilers or language extensions such as NUMBA \cite{Lam:2015:NLP:2833157.2833162} or Cython \cite{behnel2010cython}, or coded in C/Fortran and interfaced to Python.

\paragraph{Supporting external components:}
Using Python makes it simple to load theory or likelihood codes from other Python packages. \cobaya supports direct use of theory and likelihood classes defined in other packages simply by referencing their fully-qualified name in the package, making it straightforward for modellers or experimental collaborations to release their own code packages that can be used directly. External likelihood codes can either inherit directly from \cobaya\ classes (which also support direct instantiation and use outside of Cobaya), or include a separate compatible wrapping class that can be used with \cobaya.

More information on design choices can be found in the documentation at \url{https://cobaya.readthedocs.io}.

%%%%%%%%%%%%%%%%%%%%%%%%%%%%%%%%%%%%%%%%%%%%%%%%%%%%%%%%%%%%%%%%%%%%%%%%%%%%%%%%%%%
\section{Structure of the code}\label{sec:structure}
%%%%%%%%%%%%%%%%%%%%%%%%%%%%%%%%%%%%%%%%%%%%%%%%%%%%%%%%%%%%%%%%%%%%%%%%%%%%%%%%%%%

\begin{figure}[ht]
%  \centering
  \includegraphics[width=0.8\columnwidth]{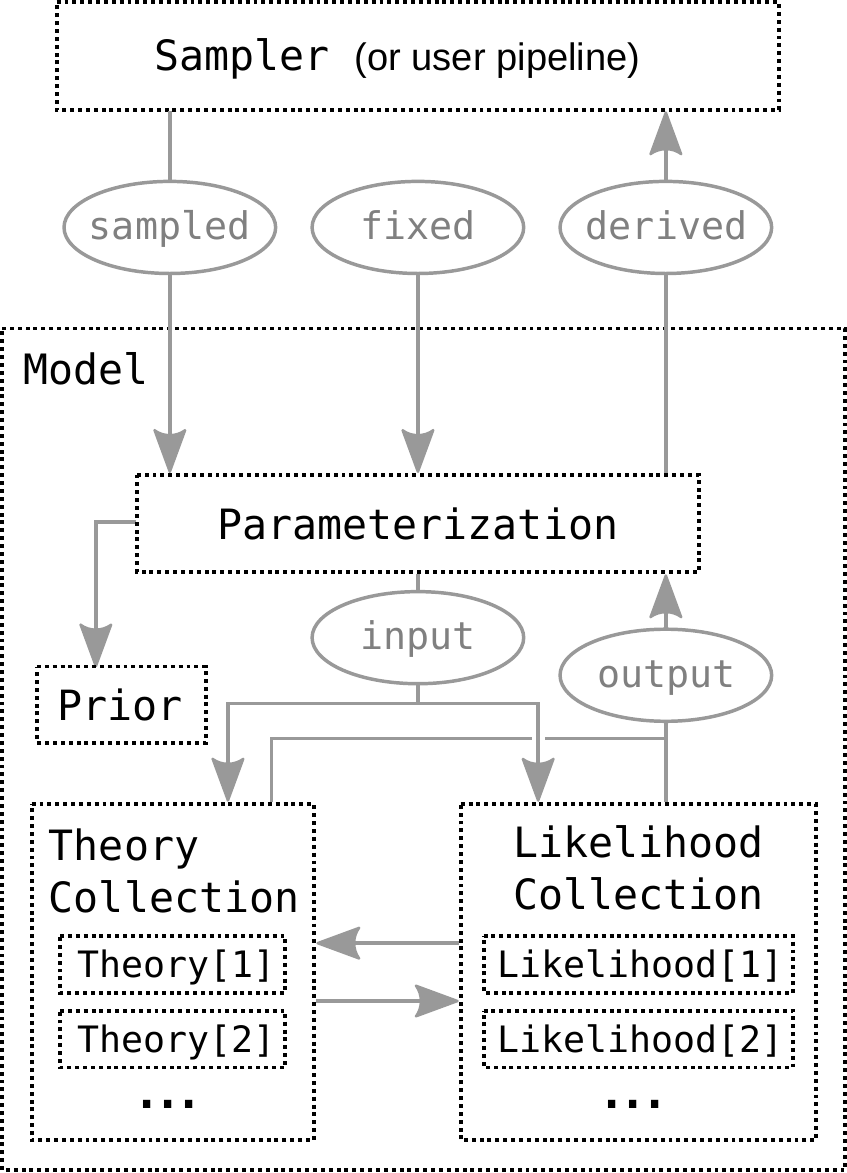}
  \caption{Simplified structure of \cobaya's source, showing classes (squares) and parameters (ellipses). See section \ref{sec:structure} for a description of each class and parameter role. The arrows between \texttt{TheoryCollection} and \texttt{LikelihoodCollection} represent computed quantities and parameters that can be exchanged arbitrarily between theories and likelihoods.
  }
  \label{fig:structure}
\end{figure}

\cobaya's main structure is shown in Fig.~\ref{fig:structure}. The main two classes are the Bayesian \texttt{Model}, and the Monte Carlo \texttt{Sampler} (or, more broadly, any analysis tool that operates on the model). This section describes the elements shown there.

\subsection{Input}

For each particular run, the input must describe the model and its parameter space in enough detail, and also specify the analysis tool that will be used. Ideally, the syntax of this input would reproduce the structure of the code as closely as possible (it makes it easier to swap different input chunks), and should be as literate (self-descriptive), human-readable and easy to remember as possible.

\cobaya's input takes the form of a \Python dictionary, and can be serialized in plain text in the YAML format \cite{YAML} (as long as its elements can be serialized in plain text). We show an example in Fig.~\ref{fig:input}, including specification of parameter roles and dynamic parameter definition, as well as definition of the likelihood and prior for the simple case where everything is just a basic Python function.

\begin{figure}[ht]
  \centering
\begin{lstlisting}[language=yaml, frame=single]
params:
  x:
    prior: {dist: uniform, min: 0, max: 2}
    ref: {dist: norm, loc: 0.75, scale: 0.1}
    proposal: 0.01
  y:
    prior: [0, 2]
    ref: {dist: norm, loc: 0.75, scale: 0.1}
    proposal: 0.01
  # Derived parameters, as function of sampled
  r:
    derived: "lambda x,y: np.sqrt(x**2 + y**2)"
  theta:
    derived: "lambda x,y: np.arctan(y/x)"
    latex: \theta

prior:
  x_eq_y_band: "lambda x,y: stats.norm.logpdf(
                   x - y, loc=0, scale=0.3)"

likelihood:
  ring: "lambda x, y: stats.norm.logpdf(
            np.sqrt(x**2 + y**2),
            loc=1, scale=0.02)"

sampler:
  mcmc:
    Rminus1_stop: 0.001

output: chains/ring
\end{lstlisting}
  \caption{Example input in plain text (YAML). It defines a Gaussian-ring likelihood with radius $1$ and standard deviation $0.02$, over the combination of a uniform prior $(x,y)\in(0,2)^2$ (notice the two possible different specifications used for \texttt{x} and \texttt{y}) and a Gaussian prior of standard deviation $0.3$ along the $x=y$ direction (n.b.: simple 1D priors are defined in \texttt{params}, while multidimensional ones are defined in \texttt{prior}). The likelihood, the multidimensional prior and the derived parameters \texttt{r} and \texttt{theta} are given as \Python functions (here source strings, but can be assigned \Python functions directly when working in a \Python file or shell -- for source strings \texttt{scipy.stats} and \texttt{numpy} are pre-imported as \texttt{stats} and \texttt{np} resp.). The results of this MCMC sample will be written in a folder called \texttt{chains} with file name prefix \texttt{ring}, as per the \texttt{output} option. The resulting densities can be seen in Fig.~\ref{fig:results}.}
  \label{fig:input}
\end{figure}

Describing complex models and pipelines can take a large amount of input. To facilitate these cases, we have implemented an inheritance system by which defaults for every class are automatically loaded. Options only need to be specified when their values are different from the default ones. In the example of Fig.\ \ref{fig:input}, for the MCMC sampler only the stopping criterion is specified, while the rest of the options are either inherited from the defaults of this sampler (which can be retrieved via a shell script ``\texttt{cobaya-doc mcmc}''), or automated from other options (e.g.\ a diagonal proposal covariance matrix constructed from the parameters' \texttt{proposal} property).

\subsection{Bayesian \texttt{Model}}

The Bayesian \texttt{Model} consists of the Bayesian prior and likelihood (this last one including theory and experimental likelihood components), as described in section \ref{sec:probmodel}. It also contains a \texttt{Parameterization} layer that manages the flow of parameters to and from the likelihood and prior, and computes the dynamically-defined ones, and a \texttt{Provider} that handles the exchange of parameters and computed quantities between different theory and likelihood components.

A \texttt{Model} instance can be passed as an argument to a sampler, or it can be integrated by the user into an external pipeline (using its API to access the member classes shown in Fig.\ \ref{fig:structure} and described below).

\subsubsection{Parameterization}

The \texttt{Parameterization} class controls the flow of parameters into and out of the \texttt{Model}, taking into account their roles with respect to different parts of the code.

On their way out of the model (i.e.\ from the point of view of the \texttt{Sampler}), parameters can play three different roles:
\begin{itemize}
\item \emph{Sampled} parameters are the ones whose value is to be varied and explored by the \texttt{Sampler} or the user-defined pipeline. They are identified in the input by having a defined prior.
\item \emph{Fixed} parameters are those whose value is not going to change, and are needed as input by some piece of the \texttt{Model}.
\item \emph{Derived} parameters are arbitrary functions of the rest of the parameters at every step, and are tracked and stored for the user's convenience. Functions defining them can be provided \emph{on the fly} (like \texttt{r} and \texttt{theta} in the example in Fig.~\ref{fig:input}) or can be implicitly defined inside the code of a theory or likelihood.
\end{itemize}

The \texttt{Parameterization} class processes the parameters to turn them into \emph{input} parameters for the likelihoods and theories, and requests from them the \emph{output} parameters that are needed to compute \emph{derived} parameters that cannot be computed directly from inputs.

The parameterization layer also manages other properties of the parameters, such as their labels (used for plots).

\subsubsection{Prior}

Priors for the sampled parameters can be specified in two different ways:
\begin{itemize}
\item The \texttt{prior} keyword for each parameter defines a separable product of 1D priors. They are interfaced directly from the 1D continuous distributions of \texttt{scipy.stats}\footnote{\url{https://docs.scipy.org/doc/scipy/reference/stats.html\#continuous-distributions}}. In the example in Fig.\ \ref{fig:input}, these are the \emph{uniform} priors $(x,y)\in(0,2)^2$.
\item Additional, possibly-multi-dimensional priors can be defined under the global \texttt{prior} block. In the example in Fig.\ \ref{fig:input}, this is the Gaussian prior along $x=y$.
\end{itemize}

\subsubsection{Likelihood and theory}
\label{sec:liketheo}

\begin{figure*}[ht]
  \centering
  \begin{subfigure}{0.5\textwidth}
\begin{python}
import numpy as np
import scipy.stats as stats

from cobaya.theory import Theory
from cobaya.likelihood import Likelihood


class ToPolarCoords(Theory):
    params = {'x': None, 'y': None}

    def get_can_provide(self):
        return ['r', 'theta']

    def calculate(self, state, want_derived=True,
                  **params_values):
        x = params_values['x']
        y = params_values['y']
        state['r'] = np.sqrt(x**2 + y**2)
        state['theta'] = np.arctan(y / x)


class GaussianRingLikelihood(Likelihood):
    params = {'mean_radius': None,
              'std_radius': None}

    def get_requirements(self):
        return ['r']

    def logp(self, **params_values):
        r = self.provider.get_result('r')
        mean_radius = \
            params_values['mean_radius']
        std_radius = params_values['std_radius']
        return stats.norm.logpdf(
            r, loc=mean_radius, scale=std_radius)
\end{python}
 %   \lstinputlisting[language=python, frame=single, linewidth=0.99\linewidth]{src_classes.py}
  \end{subfigure}\hfill
  \begin{subfigure}{0.48\textwidth}
\begin{lstlisting}[language=yaml, frame=single, linewidth=0.99\linewidth]
params:
  x:
    prior: [0, 2]
    ref: {dist: norm, loc: 0.5, scale: 0.01}
    proposal: 0.01
  y:
    prior: [0, 2]
    ref: {dist: norm, loc: 0.5, scale: 0.01}
    proposal: 0.01
  # Likelihood nuisance parameters
  mean_radius:
    prior: {dist: norm, loc: 1, scale: 0.001}
  std_radius: 0.02

prior:
  x_eq_y_band: "lambda x, y: stats.norm.logpdf(
                   x - y, loc=0, scale=0.3)"

theory:
  src_classes.ToPolarCoords:

likelihood:
  src_classes.GaussianRingLikelihood:

sampler:
  mcmc:
    Rminus1_stop: 0.001

output: chains/ring
\end{lstlisting}
  \end{subfigure}
  \caption{
    Example similar to the one in Fig.\ \ref{fig:input}, now using \cobaya classes to split the computation in two: the transformation between orthogonal and polar coordinates, and the likelihood in terms of polar coordinates. Here we let the mean radius of the gaussian ring vary over a narrow prior, to illustrate \cobaya's automated blocking: when sampling using MCMC or \polychord, jumps in the $(x, y)$ directions will be alternated with jumps in \texttt{mean\_radius}. After every jump on $(x, y)$, the resulting intermediate product $r$ is cached, so that $r$ does not need to be recomputed when only \texttt{mean\_radius} is varied. In this trivial example, the intermediate quantity $r$ exchanged between the theory and the likelihood is just a real number, but it could be any arbitrarily complicated and many-dimensional numerical quantity, as well as a general Python object. \cobaya knows about the interdependency between the likelihood, that needs $r$, and the theory, that computes $r$, via the respective declarations in methods \texttt{Likelihood.get\_requirements} and \texttt{Theory.get\_can\_provide}.}
  \label{fig:external}
\end{figure*}

In general there can be multiple theory and likelihood components in a single model, with each theory component calculating some quantity required as input to the likelihoods or to another theory component. Likelihoods are just a special subclass of a general Theory class that directly return likelihood values. Both inherit a caching layer that increases efficiency when steps in the parameter space are blocked in particular ways, such that only the piece that depends on the varied parameters (directly or via a dependency) needs to be recomputed.

A model class instance holds a list of the required likelihood and theory component instances, and uses their interface methods or introspection to work out their dependencies and hence required execution order in order to calculate the final likelihood. Components may also calculate derived parameters that are useful to include in the output parameter chain (or may be used by other components).

In  \cobaya's input, likelihoods and theory components can be specified as \Python functions, mentions to \emph{internal} classes (i.e.\ those distributed with \cobaya, either as stand-alone versions or as wrappers to external code/data), inherited classes, or mentions to classes distributed in external packages.

A simple example illustrating the use of \cobaya Theory and Likelihood classes can be seen in Fig.\ \ref{fig:external}. This example highlights one of the biggest advantages of \cobaya's Bayesian model API with respect to most generalist statistical languages: the specification of what is needed and what is provided by each stage of the pipeline is enough to produce effective posterior exploration. Automatically, \cobaya will separate the sampled parameters in two blocks according to their dependencies, [\texttt{x}, \texttt{y}] and [\texttt{mean\_radius}], it will sort them optimally (in this simple case [\texttt{mean\_radius}] comes obviously last), it will prompt the sampler, either MCMC of \polychord, to vary the parameters block-wise, and it will make sure that the intermediate quantity \texttt{r} is cached so that variations of \texttt{mean\_radius} only do not require its recomputation.

\subsection{Sampler}

Monte Carlo samplers in \cobaya take models and explore their \emph{sampled} parameters.

\cobaya implements adaptive fast-slow-optimized MCMC samplers, translated from \cosmomc \cite{Lewis:2002ah,Lewis:2013hha}. This includes a Metropolis-Hastings MCMC \cite{Metropolis53,Hastings70} sampler, with optional oversampling of fast parameters, and a variation of it that performs \emph{dragging} \cite{Neal04} along the fastest directions. In many physical applications, complex interdependencies between multiple theory codes and experimental likelihoods produce a slow-fast parameter hierarchy. Efficient exploitation of this hierarchy, while keeping the ability to propose steps in random directions of the parameter space, is crucial to efficient sampling \cite{Lewis:2002ah, Lewis:2013hha}.

Previous implementations have required a manual specification of the parameter blocking and oversampling configuration. For \cobaya\ we have implemented a new automated optimization scheme that blocks parameters according to their role in the different stages of the likelihood, measures the computational cost of varying parameters in each of these blocks, and determines the optimal parameter block sorting so that varying all parameters in a row takes the smallest amount of time possible, taking into account both parameter mixing (necessary to explore possible degeneracies) and the amount of oversampling requested by the user. For details on this algorithm see Appendix \ref{app:blocking}.

Chain sampling convergence is quantified using a modified version of the Gelman-Rubin $R-1$ statistic \cite{Gelman92,Brooks98}, either over several chains run in parallel, or over the latest chunks of a single chain. The $R-1$ statistic quantifies the variance in the means of parameters estimated from different chains (or chain subsets), and termination of the chain can be specified by giving a target small value for $R-1$. Where tail exploration is important, an equivalent stopping target can also be set for the dispersion of confidence limits.

In addition, \cobaya contains a wrapper for the nested sampler \polychord \cite{2015MNRAS.450L..61H,2015MNRAS.453.4384H}, which can also estimate model evidences and explore complicated multi-modal likelihood surfaces. \polychord\ can also exploit the parameter speed hierarchy determined by \cobaya's parameter blocking. \cobaya provides an installer for \polychord, as well as for all supported external dependencies. Wrappers for additional samplers may be implemented in the future.

Under the \textit{sampler} category, \cobaya also includes interfaces to minimizers (those in \scipy \cite{scipy} and the \pybobyqa implementation \cite{2018arXiv180400154C} of the \bobyqa algorithm \cite{Powell2009}), a simple test function to evaluate different quantities of the model at fixed points, and an importance-reweighting tool.

\subsection{Analysis -- interface to \getdist}

\cobaya manages Monte Carlo samples as wrapped \texttt{DataFrame} objects from \pandas~\cite{mckinney-proc-scipy-2010}. When written to the hard drive, they can be stored in plain text as parameter tables, including the corresponding probabilities and sample weights. Both the sample objects and output files can be easily loaded and analysed with the user's tool of choice.

\begin{figure}[ht]
  \centering
  \includegraphics[width=0.9\columnwidth]{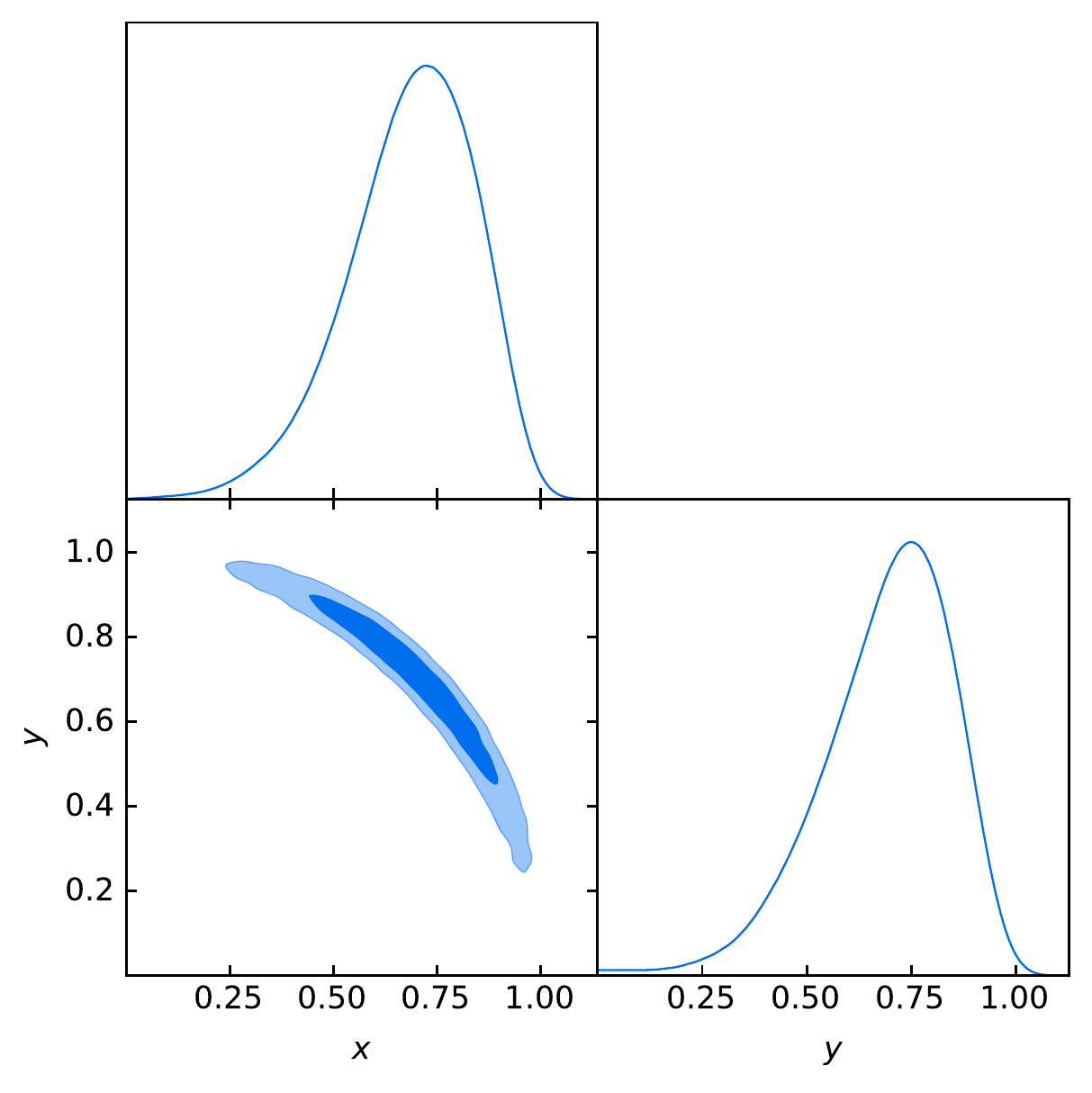}
  \includegraphics[width=0.9\columnwidth]{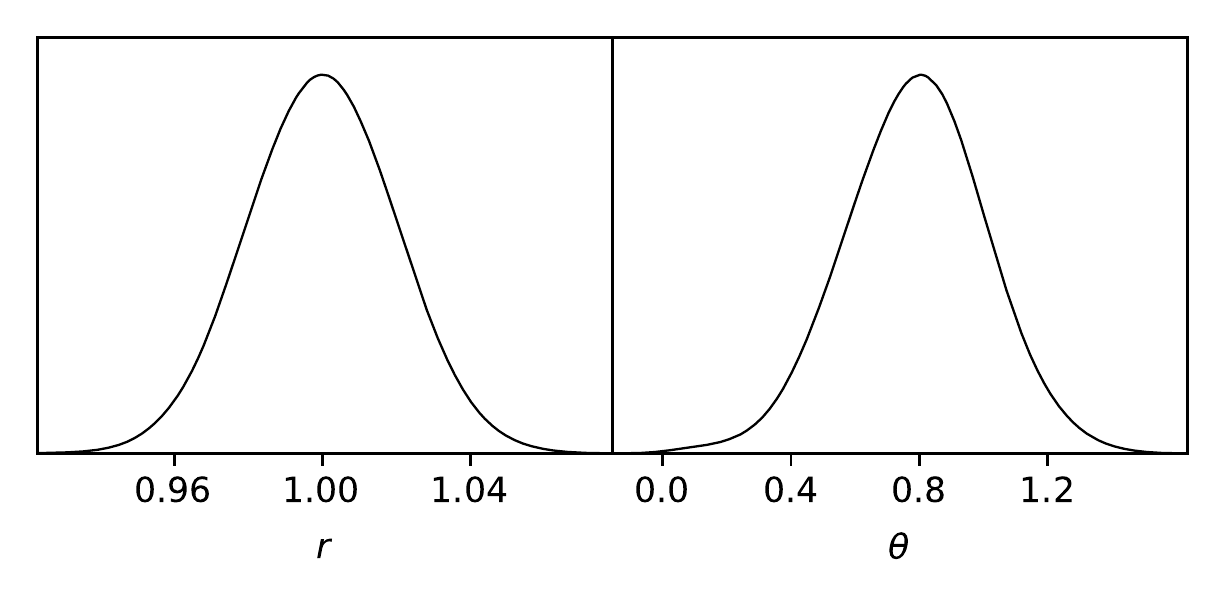}
  \caption{Results from sampling using the inputs shown in Fig.\ \ref{fig:input}, analysed by \getdist: The upper figure shows a \emph{triangle plot} combining 2D posterior contours (enclosing 68\% and 95\% of the probability) and marginalized 1D posteriors for the sampled parameters $(x,y)$. The lower plot shows the 1D posteriors for the derived parameters $(r,\theta)$. All the posteriors are shown normalized to the same maximum.}
  \label{fig:results}
\end{figure}

The suggested analysis package is \getdist\footnote{\url{https://github.com/cmbant/getdist/}} \cite{Lewis:2019xzd}, which can load \cobaya results transparently. \getdist provides summary statistics including confidence intervals, density estimates (via optimized kernel density estimation), and convergence diagnostics, plotting tools, and a graphical user interface. Examples of \getdist outputs can be seen in Fig.~\ref{fig:results}, obtained from the model described by the input in Fig.~\ref{fig:input}.

%%%%%%%%%%%%%%%%%%%%%%%%%%%%%%%%%%%%%%%%%%%%%%%%%%%%%%%%%%%%%%%%%%%%%%%%%%%%%%%%%%%
\section{High performance computing with \cobaya}
\label{sec:hpc}
%%%%%%%%%%%%%%%%%%%%%%%%%%%%%%%%%%%%%%%%%%%%%%%%%%%%%%%%%%%%%%%%%%%%%%%%%%%%%%%%%%%

\cobaya is tailored for high-performance computing on clusters. The overhead of \cobaya per posterior evaluation is $\alt 0.2\,\mathrm{ms}$, with no strong dependence on the dimensionality for $d<100$; this means that for theory or likelihood computations that take more than a few milliseconds to compute, any number of evaluations while sampling their posterior in \cobaya would not take significantly longer than evaluating the same code in a simple loop the same number of times. For higher dimensions, the components of the overhead that have a stronger dependence on dimensionality start taking over, but usually so does computation time of likelihoods that depend on an increasingly high number of parameters.

\subsection{Parallelization}

\cobaya takes advantage of hybrid MPI+threading parallelization. All samplers are MPI-aware and use communication between parallel processes to improve sampling efficiency (e.g.\ MCMC uses all chains to learn a better covariance matrix and to assess convergence). Each single MPI process carries an independent \texttt{Model}, which takes advantage of threading to accelerate its likelihood and theory computations. Some of the sampler-intrinsic operations can also take advantage of threading (e.g.\ recomputing the proposal covariance matrix). For most sampling algorithms MPI usage is a very small fraction of the computing time, so interconnect speed is largely irrelevant and the code will run well on cheap commodity (or virtualized) clusters.

Threading is usually leveraged via \numpy's \cite{numpy} interface to a threading-aware linear algebra library such as \openblas\footnote{\url{https://www.openblas.net/}}, or via the use of externally linked compiled codes where necessary.

\subsection{Batch runs}

In physical applications, it is usual to test a number of theoretical models against different combinations of data sets. Running the sampler for every combination in this \texttt{grid} of theory--data requires a large amount of very redundant input: it differs little between cases in the grid.

In order to facilitate that, we have incorporated a piece of code first implemented for \cosmomc, that can define such a \texttt{grid} starting from common definitions plus two lists of variations: theoretical models and data sets. This code prepares input files containing all possible (or allowed) combinations, creates a set of nested \texttt{model/data\_set} folder structures to store the results, and generates scripts to be submitted to a cluster queue.

\subsection{Cloud computing}

Users without access to a cluster may be able to pay for on-demand online computing resources, or cloud computing. We have explored and documented\footnote{\url{https://cobaya.readthedocs.io/en/latest/cluster_amazon.html}} \cobaya's usage in one popular solution, Amazon's EC2.

%%%%%%%%%%%%%%%%%%%%%%%%%%%%%%%%%%%%%%%%%%%%%%%%%%%%%%%%%%%%%%%%%%%%%%%%%%%%%%%%%%%
\section{Cosmology}
\label{sec:cosmo}
%%%%%%%%%%%%%%%%%%%%%%%%%%%%%%%%%%%%%%%%%%%%%%%%%%%%%%%%%%%%%%%%%%%%%%%%%%%%%%%%%%%

Although \cobaya\ is a general analysis tool, its development was originally motivated for its use in cosmology, using various standard codes for calculating cosmological observables and a variety of publicly available likelihoods from different data sets. We have included tools specific to this use case and have used it for testing on real-world problems. The use of any sampler in a cosmological context presents a series of challenges, some already presented in section \ref{sec:goals}, and others intrinsic to Cosmology:
\\

\textbf{Input complexity:} To deal with the large amount of information needed to describe cosmological models and data sets, we have created a graphical application to generate input based on the user's choice for each piece of the cosmological model (including usual priors in the literature), and the data sets used. Also, this tool automatically selects an optimal sampler configuration for each particular problem (including picking an appropriate proposal covariance matrix for MCMC from a database of standard cosmological runs).
\\

\textbf{External dependencies:}
In general, we do not modify and re-distribute existing publicly available theory and likelihood codes, but instead interface stock versions through light wrappers\footnote{We do re-implement a few dependencies (e.g.\ the $H_0$, BAO and DES cosmological likelihoods), specially when the cost of interfacing very simple external code outweighs that of maintaining our own version, or because of other potential benefits, such as extensibility, re-usability for other experimental pipelines, etc.}. This allows us to keep \cobaya's source light and more easily maintainable, letting us focus on the statistical and user experience aspects. We mitigate the added complication of having to download external code by providing a one-line installer for all those external packages, that works across multiple systems (GNU/Linux, macOS and Windows, subject to OS compatibility of each of those external codes). The installer script can take an input file and install everything necessary to run it, including theory and likelihood codes, and data sets.
\\

\textbf{User-modified external code:} By not repackaging and re-distributing our own version of external codes, we impose no entry barrier for user-modified versions: users just need to provide paths to the installation folders of their modified versions.
\\

\textbf{Alternative external codes with the same role:} We have made the interface between experimental likelihoods and cosmological theory codes agnostic to the theory code used (as long as the quantities requested by other parts of the code can be computed by all of the alternatives).
\\

\textbf{Computational cost:} For cosmological data probing non-background cosmology, a likelihood evaluation for given cosmological parameters typically requires running at least a Boltzmann code to calculate the perturbation transfer functions (taking a second or more depending on number of cores and the specific model), plus possibly additional steps such as modelling non-linear evolution, calculating angular power spectra, correlation functions, etc.\ from the transfer functions. The second step can be fast in linear theory, or seconds when non-linear modelling is involved or complex observables. The original \cosmomc\ code provided a split between cosmological parameters affecting the transfer functions and parameters governing in the initial power spectrum, allowing the latter to be sampled very quickly for fixed values of the transfer function parameters~\cite{Lewis:2002ah}, but assumed linear theory. In \cobaya\ the parameter dependencies and possible speed hierarchy can also be exploited by splitting the theory calculation up into a transfer function calculation, and subsequent calculations involving initial power spectrum parameters and non-linear evolution parameters. \cobaya's \camb\ wrapper explicitly supports this with non-linear modelling based on variations of \textsc{halofit}~\cite{Smith:2002dz,Takahashi:2012em,Mead:2015yca}, generalizing the approach used in \cosmomc. Although the speed hierarchy is modest once non-linear corrections to the CMB lensing\footnote{A Limber approximation for the non-linear corrections could likely speed this up with very small loss of accuracy, but is not currently implemented.} and other spectra are included, the saving can be significant since at the very least the cost of computing the transfer functions does not have to be paid every time power spectrum and non-linear evolution parameters are varied.

Despite the small additional Python overhead, \cobaya cosmology MCMC chains runs using the $\Lambda$CDM model with \camb and {\it Planck} data~\cite{PCP2018}  have comparable computation cost to using \cosmomc. In particular, assuming a good starting guess of the proposal covariance matrix,\footnote{This is done to reduce random noise in the test, by removing the very variable initial fraction of samples during which most of the proposal covariance matrix is learned. Also notice that in \cosmomc (and also in \montepython and \cosmosis at the moment of writing) the proposal covariance matrix guess needs to be given by hand, while in Cobaya it is automatically selected from a database for a range of cosmological configurations.} both \cobaya and \cosmomc consistently reach Gelman-Rubin convergence of $R-1\alt 0.02$ in 8 hours on four cores per chain and 4 chains. Equivalence with \cosmomc as the \textit{de facto} standard in the field should be enough, specially when considering the additional advantages of \cobaya. In particular, \cobaya's model API (see section \ref{sec:liketheo}) and its optimal-sorting algorithm (see Appendix \ref{app:blocking}) guarantee that as pipelines become more complex and are further modularised, \cobaya will consistently outperform simpler approaches.

For example, \cobaya\ would be more efficient when varying a larger number of primordial power spectrum parameters with non-linear modelling, e.g.\ when performing primordial power spectrum reconstruction (which is explicitly supported in \cobaya\ by using a separate primordial power spectrum Theory class; currently compatible with \camb only). For application to future data the calculation could be further modularized. For example, the non-linear modelling could be separated into a separate Theory class taking inputs from a Boltzmann code and calculating non-linear power spectra. This could be implemented in different ways and used consistently by likelihoods using different class implementations for cross-comparison. The calculation of actual observables from the underlying power spectra, for example by codes such as CCL\cite{Chisari:2018vrw}, could be implemented as a separate Theory class taking non-linear power spectra and  outputs from a Boltzmann Theory code, and using them to calculate observables such as tomographic lensing correlation functions required as the theoretical model input used by likelihoods.

%%%%%%%%%%%%%%%%%%%%%%%%%%%%%%%%%%%%%%%%%%%%%%%%%%%%%%%%%%%%%%%%%%%%%%%%%%%%%%%%%%%
\subsection{An example use case in Cosmology}
\label{sec:cosmo_use_case}
%%%%%%%%%%%%%%%%%%%%%%%%%%%%%%%%%%%%%%%%%%%%%%%%%%%%%%%%%%%%%%%%%%%%%%%%%%%%%%%%%%%

% Cosmo framework
As a demonstration of some of \cobaya's capabilities when applied to a cosmological scenario, we present here an inference problem in which we search for a primordial power spectrum feature of the form
\begin{equation}
  \label{eq:feature}
  % Delta P/P = A * exp(-0.5 (log10(k/c)/w)**2) * sin(2 pi k/l)
  \frac{\Delta P(k)}{P_0(k)} = A \exp^{-0.5 \left[\frac{\log_{10}\left(\frac{k}{k_c}\right)}{w}\right]^2} \sin\left(2 \pi \frac{k}{l}\right)
  \,,
\end{equation}
where $P_0(k)$ is a power-law nearly-scale-invariant scalar power spectrum, that has been injected into a fiducial cosmological model. This ansatz has four parameters: a relative \emph{amplitude} $A$ of the feature, a \emph{wavelength} $l$, and an \emph{envelope centre} $k_c$ and \emph{envelope log-width} $w$. We forecast how well the feature can be characterized using highly idealized versions of the Planck and forthcoming Simons Observatory (SO) data~\cite{Ade:2018sbj,PL2018}. To use the maximal effective sky area at each multipole we separate the likelihood into low- and high-$\ell$ ranges, with different sensitivities, sky fractions, and polarization combinations (we will see that a single likelihood wrapper can incorporate all of these).

We try to constrain/recover the injected feature parameters in two scenarios: one using standard lensed CMB power spectra, and in which we assume that a certain amount of delensing~\cite{Sherwin:2015baa,Larsen:2016wpa,Green:2016cjr} is possible. Large-scale lensing modes shift the scales of small-scale oscillation modes leading to them being smoothed out in the lensed power spectrum. If lensing multipoles $L\alt 200$ can recovered using an external tracer like the Cosmic Infrared Background (CIB) or internally reconstructed, this smoothing effect can be substantially reduced by delensing. As a simplified model we simply assume that the lensing power can be reduced by a constant factor of $0.7$ using a combination of internal delensing, CIB, and large-scale structure\footnote{The detectability of small-scale features is not very sensitive to whether or not the $L>200$ lensing modes are also delensed, since they largely contribute to adding power to both spectra.}. We do not attempt to model the statistics of the delensed power spectrum in detail, and take the log likelihood of both delensed and lensed spectra to be given by an $f_{\rm sky}$-scaled ideal full-sky mean log likelihood.

We demonstrate here how \cobaya can be used to build a pipeline for this use case. To do this, we take advantage of \cobaya's \camb interface for computing the CMB power spectrum given a primordial spectrum, that we will define in a separate Theory class; we will also create a generic Likelihood class using data that we have generated based on a fiducial CMB power spectrum containing a feature. The step-by-step building of this pipeline, would proceed in the following way (the source code of a possible implementation is included in Appendix \ref{sec:cosmo_demo_source}):
\begin{itemize}

\item We define a Theory class that computes the primordial power spectrum including the feature (see \texttt{FeaturePrimordialPk} in Fig.\ \ref{fig:src_demo_theory}). This class, inheriting from the \cobaya generic Theory prototype class, only needs to define two methods: one returning what this class can compute and starting with \texttt{get\_}, in this case \texttt{get\_primordial\_scalar\_pk}; and another called \texttt{calculate} taking the parameters defining the power spectrum and computing it, in the example of Fig.\ \ref{fig:src_demo_theory}, simply a wrapper around a Python function that defines a linearly-oscillating feature with a Gaussian envelope on top of a nearly-scale-invariant power spectrum. The \emph{class attribute} \texttt{params} defines what parameters among all the ones declared in the input are to be passed to this class (and also which derived parameters are made available by this likelihood). The rest of the class attributes (in our example \texttt{n\_samples\_wavelength} and \texttt{k\_pivot}), define parameters that can be set in the declaration of the class in the \cobaya input (see Fig.\ \ref{fig:src_demo_input_delensed}), and their default values if they are omitted.

\item Regardless of the definition of the parameters passed to the new Theory class, we may want to redefine them e.g.\ to sample them over a log-uniform prior. We show how to do that for the feature \texttt{amplitude}, its \texttt{wavelength} and the \texttt{centre} of its envelope in Fig.\ \ref{fig:src_demo_input_delensed}: we define the respective dummy log-parameters in the \texttt{params} block, indicate that they shall not be passed to the pipeline with the option ``drop: True'', and define the original ones as a function of them. These redefinitions are specially useful when we cannot easily modify the source code of the original Theory code or likelihood (which is a common occurrence for complex codes such as official experimental likelihoods).

\item We may want to impose additional priors that depend on more than one parameter. In our example, we would like to ban combinations of amplitude and position and width of the feature envelope that produce no significant trace in the observable CMB multipole range. To do that, we can define the corresponding log-prior function e.g.\ in the same file as the Theory code (see \texttt{logprior\_high\_k} in Fig.\ \ref{fig:src_demo_theory}), and reference it in the input as shown in Fig.\ \ref{fig:src_demo_input_delensed}.

\item The likelihood segment of the pipeline is usually the most complicated one for research cases that do not use public data as-is, or public likelihood codes. Users can take two alternative approaches: on the one hand, they can write generic (i.e.\ non-\cobaya-dependent) code to generate or load some data and compute their likelihood (possibly assuming a parameterized experimental model), and then wrap it in a \cobaya Likelihood class; alternatively they can inherit from one of \cobaya's cosmological likelihoods and override the necessary methods. In both cases, the methods needing defining/overriding are \texttt{get\_requirements}, declaring which cosmological observables/quantities will be requested from the Theory code(s), and \texttt{logp}, that retrieves said quantities and passes them to the user-defined probability function (or does this computation directly). These two methods for our use case can be seen in Fig.\ \ref{fig:src_demo_likelihood}. Here, we inherit from one of \cobaya's likelihoods, \texttt{CMBlikes}, that takes care of loading the necessary data and computing a mean log-likelihood given the polarized CMB power spectra. We override \texttt{get\_requirements}  to add the \emph{results} object of \camb (\texttt{CAMBdata}, from which we will extract the partially-lensed CMB spectra), to the list of CMB polarizations and multipole ranges defined by the data with which the likelihood was initialized (and set here via a \texttt{super} call). This shows the flexibility of \cobaya's wrappers: even if \cobaya does not know about a particular observable, it can still be retrieved from the theory code by hand. We also need to override the \texttt{logp} method of this class to retrieve the partially-lensed power spectra instead of the fully-lensed ones.

Notice that likelihoods inheriting from an existing cosmological \cobaya likelihood can work outside the \cobaya pipeline: they can e.g.\ be used as standalone objects that can evaluate data probabilities given an arbitrary realization of the required observable. As with the Theory codes, we make use of \emph{class attributes} to specify which options can be defined in the input at run time, which allows us to write a single likelihood wrapper for our four likelihoods, with the respective data specified via \texttt{dataset\_file}, and the lensing scenario defined by the value of \texttt{Alens\_delens} ($0.3$ for the delensed case, and $1.0$ for the non-delensed one).

\item On the sampler side, here MCMC, thanks to our use of \cobaya's class wrappers, the parameter speed hierarchy is taken advantage of automatically: parameters are split into a \emph{primordial power spectrum} block (those sent to \texttt{FeaturePrimordialPk}) and a \emph{CMB transfer functions} block (those sent to \camb), and these last ones, as they are the slowest, are placed first. Since we are exploring both feature and baseline $\Lambda$CDM parameters, and the latter will have approximately similar posteriors in delensed and non-delensed cases (since we are injecting a feature for which we don't expect strong correlations with background cosmology), we may want to perform an initial run with $\Lambda$CDM cosmology and a liberal convergence criterion, that we will use to generate a covariance matrix (a standard output file of \cobaya's MCMC sampler). We can then set this file as the proposal \texttt{covmat} for the \texttt{mcmc} sampler (see Fig.\ \ref{fig:src_demo_input_delensed}). The \texttt{mcmc} sampler will then create the total covariance matrix, using the provided one (including correlations) for the $\Lambda$CMD parameters, and the \texttt{proposal} property of each parameter for the feature parameters.
\end{itemize}

Notice that when defining this pipeline, we did not need to modify \cobaya or \camb, but simply write the code carrying out the calculations in a non-\cobaya-dependent way, and code \cobaya wrappers for these calculations, what they require, and what they make available. The advantages of this approach are on the one hand that it is more accessible for researchers that do not know the inner workings of \cobaya (or \camb), and on the other that distributing a small set of files (for reproducibility and archiving purposes\footnote{Reproducibility is ensured despite possible backwards-incompatible changes in the codes used, since \cobaya stores both its version and those of the external codes used (here \camb) in the output files.}) is more practical and scalable than distributing whole modified versions of codes.\footnote{The source code to reproduce this example can be downloaded from \url{https://github.com/CobayaSampler/paper_cosmo_demo}}

\begin{figure*}[ht]
  \centering
  \includegraphics[width=.8\textwidth]{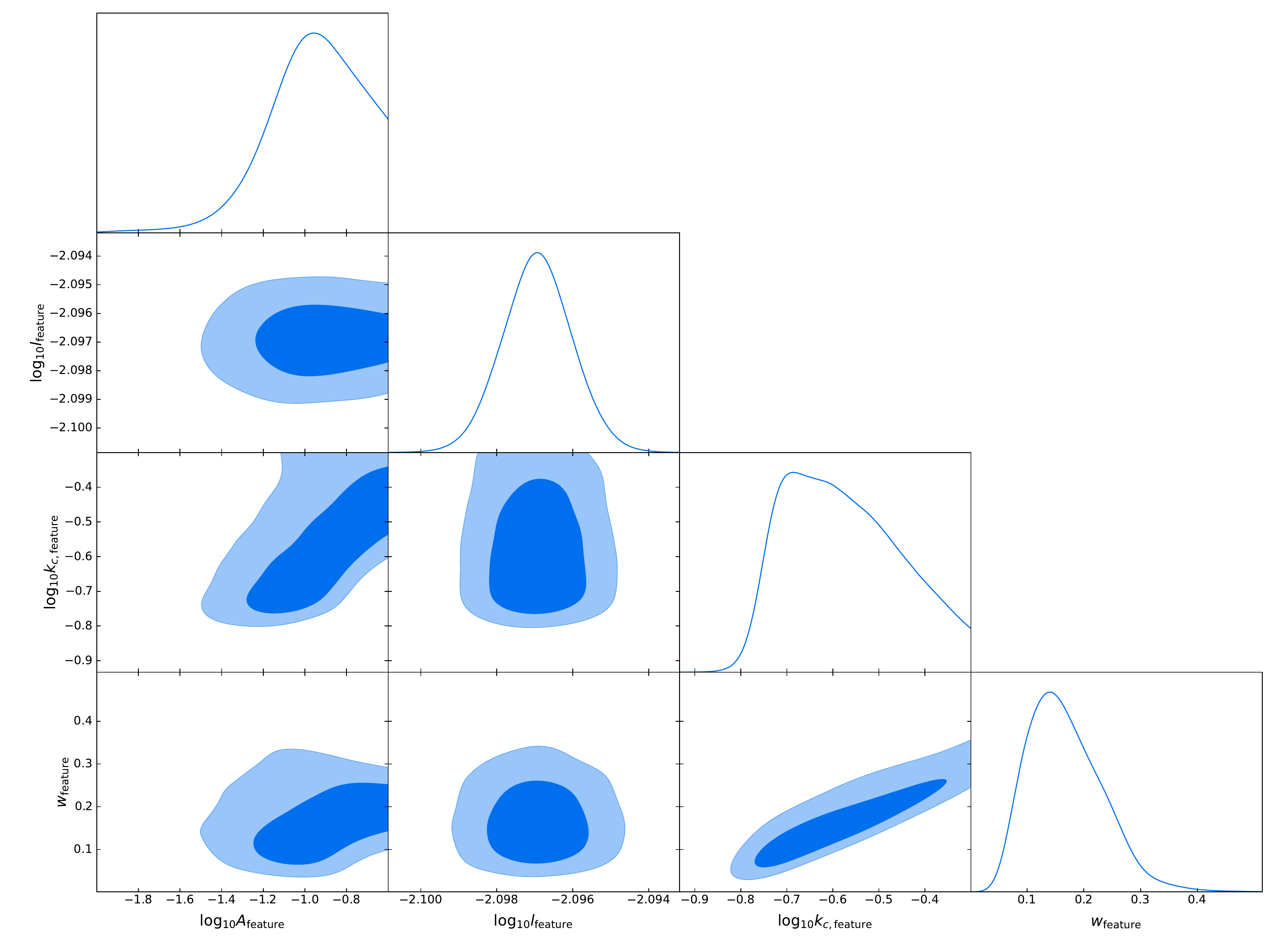}
  \caption{1- and 2-d marginalized posteriors for the feature parameters in the delensed scenario ($\Lambda$CDM cosmological parameters were sampled but are not shown here). There exists a degeneracy towards simultaneous large amplitude, wide envelope and large-$k$ center reaching the prior boundaries (see discussion in main text).}
  \label{fig:cosmo_demo}
\end{figure*}

In order to test our pipeline, we have produced both a delensed and a non-delensed MCMC run based on a fiducial cosmological model containing a feature as defined in Eq.\ \eqref{eq:feature} with fiducial parameter values $A=0.1$, $l=0.008\,\mathrm{Mpc}^{-1}$, $k_c=0.2\,\mathrm{Mpc}^{-1}$, $w=0.1\,\mathrm{Mpc}^{-1}$. For the rest of the cosmological model, we assume Planck-like $\Lambda$CDM \cite{Aghanim:2018eyx} with parameter values given by the best fit to the low-$\ell$ TT+EE, high-$\ell$ TT+TE+EE and lensing power spectrum Planck 2018 likelihoods. We vary both feature and $\Lambda$CDM parameters, with a prior for the feature parameters centered around the fiducial values (the aim for this demonstration is to characterize the injected feature, not to detect its presence). We show the result for the delensed scenario in Fig.\ \ref{fig:cosmo_demo} (for the sake of brevity, the marginalized 1- and 2-d posteriors for the rest of the cosmological parameters are not shown).\footnote{The sampling process took 72h (wall-clock time) to converge (Gelman-Rubin $R-1<0.01$) running 4 parallel chains with 4 threads each, on an Intel Xeon CPU E5-2640v3. The running time is significantly higher than in the test case discussed in Sec.\ \ref{sec:cosmo} due to the increased precision necessary to run this pipeline (at the levels of transfer function computation, CMB multipole integration, and lensing potential computation), which makes the full pipeline in this setting take $\approx 7s$.} Not surprisingly, we find a degeneracy between the feature, at the same time, (1) being centered towards higher $k$, (2) having a wider envelope, and (3) having a bigger amplitude. We cannot usually map the end of this degeneracy because it would involve considering amplitudes $\Delta P(k) / P(k)\sim 1$, which would contradict the assumption implicit in Eq.\ \ref{eq:feature} that the feature is a small perturbation on top of vanilla slow-roll dynamics producing nearly-scale-invariant power-law primordial spectrum.

In the non-delensed case, not shown in Fig.\ \ref{fig:cosmo_demo}, the MCMC chains mostly reproduce the prior region in which the feature does not leave a significant imprint on top of baseline $\Lambda$CDM (i.e.\ small feature amplitudes, envelopes centered outside the CMB window, or envelopes being too narrow). This is due to the small imprint the fiducial feature leaves on top of $\Lambda$CDM when full lensing is considered.

This example use case is presented as a demonstration of an array of \cobaya's features, and it should not be taken as proof that delensing significantly enhances primordial feature recovery. While this is probably true for a small region of the parameter space, one would need to perform a more detailed study, using more realistically-generated data and more realistic delensing, and characterizing the $\chi^2$ distribution over a number of simulations to assess detectability. The resulting posterior would also be more irregular than the smooth one shown in Fig.\ \ref{fig:cosmo_demo}, and possibly multimodal, in which case it would make sense to use \polychord instead as a sampler.

%%%%%%%%%%%%%%%%%%%%%%%%%%%%%%%%%%%%%%%%%%%%%%%%%%%%%%%%%%%%%%%%%%%%%%%%%%%%%%%%%%%
\section{Conclusion}
%%%%%%%%%%%%%%%%%%%%%%%%%%%%%%%%%%%%%%%%%%%%%%%%%%%%%%%%%%%%%%%%%%%%%%%%%%%%%%%%%%%

\cobaya is a general modularized framework for Bayesian analysis of models with complex pipelines, with simple and structured input specification, and a minimal API for interfacing external theories and likelihoods. Model posteriors can be sampled, maximized and importance-reweighted. \cobaya's main advantage is its ability to automatically account for interdependencies between the different components of the pipeline, and to use these dependencies and the computational costs of varying each parameter to optimally sort the parameter blocks of the model so that the posterior is efficiently explored, all without the need for specific code by the user.

\cobaya is well suited for high-performance computing thanks to its low overhead, MPI parallelization, and batch-running of grids of jobs. In a future release, we plan to implement HPC-enabled containerization capabilities (\texttt{Docker}+\texttt{Shifter} and \texttt{Singularity}), as well as to improve the batch-running tools.

%%%%%%%%%%%%%%%%%%%%%%%%%%%%%%%%%%%%%%%%%%%%%%%%%%%%%%%%%%%%%%%%%%%%%%%%%%%%%%%%%%%
\section{Acknowledgements}
%%%%%%%%%%%%%%%%%%%%%%%%%%%%%%%%%%%%%%%%%%%%%%%%%%%%%%%%%%%%%%%%%%%%%%%%%%%%%%%%%%%

Thanks to J.\ Lesgourgues and W.\ Handley for support on interfacing \class and \polychord, respectively. Thanks to G.\ Ca\~nas Herrera, A.\ Finke, X. Garrido, S.\ Heimersheim, L.\ Hergt, M.S.\ Madhavacheril, V.\ Miranda, T.\ Morton, J.\ Zunz and many other for help debugging and patching Cobaya.

We acknowledge support from the European Research Council under the European Union's Seventh Framework Programme (FP/2007-2013) / ERC Grant Agreement No. [616170].
AL is also supported by the UK STFC grants ST/P000525/1 and ST/T000473/1.

\appendix
%%%%%%%%%%%%%%%%%%%%%%%%%%%%%%%%%%%%%%%%%%%%%%%%%%%%%%%%%%%%%%%%%%%%%%%%%%%%%%%%%%%
\section{Automatic parameter blocking for speed-hierarchy exploitation}\label{app:blocking}
%%%%%%%%%%%%%%%%%%%%%%%%%%%%%%%%%%%%%%%%%%%%%%%%%%%%%%%%%%%%%%%%%%%%%%%%%%%%%%%%%%%

In realistic physical scenarios, \emph{full} likelihoods tend to be a combination of one or more \emph{experimental likelihoods} which depend on observables and other quantities computed by one or more \emph{theoretical codes}. This interdependence inevitably induces a speed hierarchy in the parameter space, since when only certain subsets of parameters are varied, the full pipeline does not need to be recomputed (provided that intermediate quantities are cached and reused). Individual theory and likelihood calculations may also have parameters with very different numerical costs to update; for example, overall normalization or calibration parameters are often very fast to update if other parameters remain fixed.

To exploit the speed differences, we group parameters into \emph{blocks} of parameters with approximately the same speed. The reason for taking parameters to be in blocks, rather than just each parameter separately, is that for parameters of approximately the same speed it can improve exploration efficiency to take proposal steps in randomized directions within the block's (affine-transformed) parameter space; this increases robustness to misestimation of the covariance, and allows non-trivial low-dimensional dependency structures to be randomly probed. For further details of our randomized blocked proposal distribution see Ref.~\cite{Lewis:2013hha}.

We could exploit the speed hierarchy in MCMC sampling by directly taking relatively more steps in the fast blocks compared to the slow blocks. But this would lead to low sampling efficiency when there are strong correlations between fast and slow parameters. However, it was shown in Ref.~\cite{Lewis:2013hha} that sorting the parameters such that slow parameter speed blocks precede the fast ones, and applying an affine transformation consisting of a lower-triangular matrix, one can preserve most of the advantages of the speed hierarchy while allowing for steps along degeneracies between parameters in different blocks: due to the lower-triangular nature of the affine transformation, variations of the slow parameters turn into variations of mixtures of fast and slow parameters, while variations of fast parameters remain fast. In this appendix, we present an algorithm to find the optimal sorting of parameter blocks when the likelihood pipeline features a non-trivial speed hierarchy due to having multiple interdependent components.

For fairly simple unimodal distributions, one can find one such affine transformation that approximately \emph{decorrelates} the parameter space (i.e.\ maps the original parameters into a frame where they are centred, uncorrelated and equally normalized). This transformation is given by a translation to the mean of the chain (with no effect on the parameter blocking) and a rotation by the (lower-triangular) Cholesky-left matrix of the covariance matrix of the Monte Carlo samples at some point during sampling, when the chain is already centered around the mode of interest, even if not sampling very efficiently. Iterating the determination of the covariance matrix allows rapid adaption of the sampling to the shape and dimensions of the true posterior, even if it is only very poorly known in advance. As long as the process of estimating the covariance converges, or the iterations are stopped after some finite time, the samples from the latter part of the chain will be Markovian, and the samples will converge to being samples from the desired distribution~\cite{Haario01}. Once the affine transformation is well determined, sampling can proceed efficiently by taking steps along the eigendirections in parameter space, allowing large moves along nearly degenerate directions of parameter space that would otherwise be very slow to explore by a random walk\footnote{In cases where derivatives of the likelihood can be calculated Hamiltonian (HMC) and other sampling methods may be much more efficient. For the current release \cobaya\ is focussed on likelihoods where derivatives are not readily available. }. Choosing a decorrelating transformation as the lower-triangular affine transformation mentioned above significantly boosts the exploitation of the optimal block sorting presented below, but it is not mandatory.

Let us present the problem with a non-trivial example: a case in which the full likelihood consists of two experimental likelihoods, $\liken_{i=1,2}$ that take time $t_{i=1,2}$ to compute and depend on $n_{i=1,2}$ experimental model parameters $\lambda_{i=1,2}$. Let both experimental likelihoods be a function of the same observable $\obs$, which depends on $n_\mathcal{T}$ theory parameters $\theta$, and takes time $t_\mathcal{T}$ to compute. The fast parameters in this case will be those of the experimental likelihoods, since varying the parameters of the theoretical model imposes the recomputation of both the theoretical model and the likelihoods. Due to the Cholesky decorrelation, changing parameters in the ordered-first likelihood block also requires recalculating second block of likelihood parameters. So the total time it takes to vary the parameters one-by-one will depend on the order in which each of the experimental likelihoods are computed (see Eq.\ \eqref{eq:cost} below): $t_{\mathcal{T}\rightarrow 1\rightarrow2}= n_\mathcal{T}\,(t_\mathcal{T}+t_1+t_2) + n_1\,(t_1+t_2) + n_2\,t_2$, and the corresponding result exchanging 1 and 2 for $t_{\mathcal{T}\rightarrow 2\rightarrow1}$. The difference between these times amounts to $\Delta t_{(1\rightarrow 2) - (2\rightarrow 1)} = n_1\,t_2 - n_2\,t_1$, which indicates that the likelihood with most parameters and that takes the shortest time to compute should be ordered last.

In realistic physical problems, it is common to find even more complex cases, with more complicated parameter dependencies including parameters shared by several experimental likelihoods, etc. We present now an algorithm to find the optimal ordering when no information about the posterior is known, including the possibility of varying each parameter more than once per parameter cycle depending on its relative speed.

The first step in the algorithm is to figure out the dependency structure between the different components of the likelihood (experimental likelihoods and theory codes), and block parameters according to those dependencies, such that two parameters belong to the same block if varying them entails the recomputation of the same components of the likelihood. To each parameter block we will assign a \emph{footprint}, i.e. for an arbitrary ordering of the likelihood components, a vector whose elements are either 1, if the component in that position needs to be recomputed, or 0 if it does not.

A typical example in Cosmology that involves complex dependencies is putting constraints on the $\Lambda$CDM model with the Planck low-$\ell$ and high-$\ell$ data set. The theory computation of the $\Lambda$CDM observables depends on 6 parameters $\vector{\theta}=(A_\mathrm{s}, n_\mathrm{s}, H_0, \Omega_\mathrm{B}, \Omega_\mathrm{CDM}, \tau_\mathrm{reio})$. Both of Planck's data sets share a single calibration parameter $\vector{c}=(c)$, and the low- and high-$\ell$ likelihood have a number of additional \textit{nuisance} parameters, $\vector{l}$ and $\vector{h}$ respectively.\footnote{Actually, the low-$\ell$ likelihood has no additional nuisance parameters beyond calibration, but let us assume it does for the sake of making the example more general} Variations of parameters in $\vector{\theta}$ impose recomputation of the theory and both likelihoods, variations of $\vector{c}$ impose recomputation of both likelihoods (but not the theory), and variations of $\vector{l}$ ($\vector{h}$) impose just the recomputation of the low-$\ell$ (high-$\ell$) likelihood. Therefore there would be four different parameter blocks here, with the following footprints:
\begin{equation}
\begin{split}
  \label{eq:footprintexample}
  \vector{f}_{\vector{\theta}} = (1,\,1,\,1) ,\quad&
  \vector{f}_{\vector{c}} = (0,\,1,\,1) ,\\
  \vector{f}_{\vector{l}} = (0,\,1,\,0) ,\quad&
  \vector{f}_{\vector{h}} = (0,\,0,\,1)
  \eqc
\end{split}
\end{equation}
where we have used the order ($\Lambda$CDM, low-$\ell$, high-$\ell$).

The next step of the algorithm is to compute the time cost of the pipeline whenever a parameter in each of the blocks is varied. To do that, we start by constructing a vector $\vector{t}$ containing the evaluation time of the different components in the order in which they appear in the footprints. One would naively think that the cost of varying a parameter in a particular block is equal to the scalar product of $\vector{t}$ and the footprint of that block, but the mixture of parameters produced by the Cholesky decorrelation transform in general requires recomputing components that do not appear with a 1 in the footprint, depending on the position of the block with respect to the rest.

To see how downwards mixing due to Cholesky decorrelation would be represented, let us create a matrix $\vector{F}$ by stacking the footprints following some ordering $\sigma$. In our example, this would be, for some particular ordering $\sigma=(\vector{\theta},\,\vector{c},\,\vector{h},\,\vector{l})$,
\begin{equation}
  \vector{F}(\sigma) =
  \left(\begin{array}{ccc}
          1 & 1 & 1\\
          0 & 1 & 1\\
          0 & \fbox{0} & 1\\
          0 & 1 & 0
        \end{array}\right)
      \eqc
\end{equation}
where rows represent blocks and columns represent likelihoods and theory in the agreed order.

When computing the cost of varying all parameters in that order, in this particular example, taking into account downwards mixing would be as simple as turning the framed zero into a one, since parameters in $\vector{h}$ would be mixed with parameters in $\vector{l}$, and this would impose the recomputation of the low-$\ell$ likelihood at this step.

%Being the evaluation time vector $\vector{t} = (t_{\Lambda\mathrm{CDM}},\,t_{\mathrm{low}\ell},\,t_{\mathrm{high}\ell})$

One can easily convince oneself that this transformation of the footprint matrix can be generalized to the operator
\begin{equation}
  \label{eq:footprintoperator}
  \mathrm{D}(\vector{F}) = \min(1, \vector{L}^\mathrm{T}\vector{F})
  \eqc
\end{equation}
where the operator $\min$ is applied element-wise, and $\vector{L}$ is a lower-triangular square matrix filled with 1's, of dimension equals the number of parameter blocks. Writing $\vector{L}^\mathrm{T}\vector{F}$ as $(\vector{F}^\mathrm{T}\vector{L})^\mathrm{T}$ it is easy to see that the effect of $\vector{L}$ is propagating the footprints upwards, which is equivalent to mixing the parameters downwards.

We are ready now to compute the total cost of varying all parameters, once each. Let there be $N$ blocks, where block $i$ has $n_i$ parameters and footprints $\vector{f}_i$. For a given ordering $\sigma\in\mathrm{Sym}(N)$ (the set of all possible permutation of $N$ elements), for a series of likelihood components with cost $t_j$, the total time cost is given by the dot product
\begin{equation}
  \label{eq:cost}
  T(\sigma) = \vector{n}_\sigma \cdot \tilde{\vector{t}}_\sigma
  \eqc
\end{equation}
where we have defined the \emph{cumulative} per-block cost vector $\tilde{\vector{t}}\defeq \mathrm{D}(\vector{F})\,\vector{t}$. In the absence of knowledge about the posterior (i.e.\ when no oversampling is assumed a priori), the optimal parameter order is the one that it minimizes  $T(\sigma)$; i.e., $\hat{\sigma} = \operatorname*{arg\,min}_{\sigma\in\mathrm{Sym}(N)} T(\sigma)$. A unique solution is not guaranteed, but not needed either.

\begin{figure}[ht]
  % \centering
  \includegraphics[width=\columnwidth]{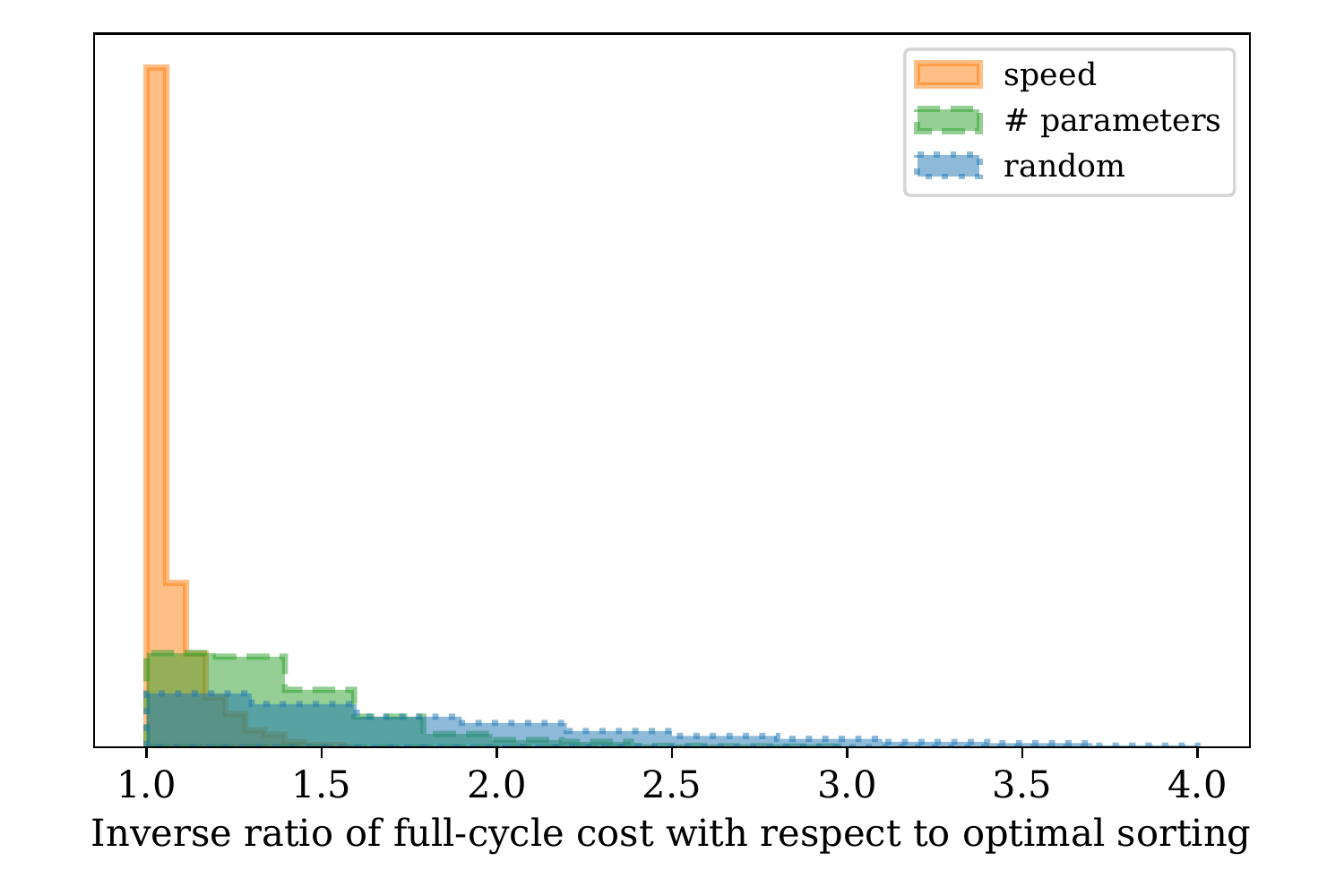}
  \caption{Performance improvement achieved by our optimal sorting over random sorting (blue, dotted line), and over sorting in terms of speeds (fastest likelihoods last; orange, continuous line) and number of parameters (likelihoods with most parameters last; green, dashed line), assuming 8 independent likelihoods with 1--20 parameters (uniformly drawn) and computation time between $10^{-2}$ and $1$ (log-uniformly drawn; no units needed).
  }
  \label{fig:optsort}
\end{figure}

To illustrate the expected gain in performance of this algorithm, we conducted a simple test assuming a number of likelihoods with no interdependencies (so that they could be computed in any order) sharing some common slow theoretical computation. We allowed these likelihoods to have different speeds and numbers of parameters on which they depend. As a test measure, we can look at the cost of cycling over all parameters consecutively, and compute the ratio between the optimal sorting provided by our algorithm, and random sorting plus two na\"ive sortings: likelihoods with more parameters last, and faster likelihoods last. In Figure \ref{fig:optsort}, we show the results for 2000 draws of sets of 8 likelihoods with input parameter numbers chosen uniformly between 1 and 20, and computation times chosen log-uniformly between $10^{-2}$ and $1$ (units are irrelevant here; just relative speeds matter). Our method easily achieves multiples of performance with respect to random sorting, and produces more modest but consistent performance gains over the two na\"ive sortings, all coming at the cost of a one-off computation taking less than a second at the beginning of the sampling. In general, for different configurations, our optimal sorting outperforms the two na\"ive sortings more clearly the more similar are the speeds and number of parameters between pipeline components. It should be noted that even small performance gains that limit the number of redundant calculations necessary will contribute to reducing the significant carbon footprint of HPC.

\subsubsection*{Oversampling of fast parameters}

Large speed hierarchies naturally arise in segmented pipelines, when experimental likelihoods need theoretical quantities to be computed externally, easily producing ratios of orders of magnitude between evaluation times of theories/likelihoods. In these cases, some oversampling of the fast parameters --- usually nuisance parameters of the experimental model --- would not add a much computational cost to the problem, while improving the quality of the posterior in those directions and reducing the overall decorrelation time. Even if there is no large speed hierarchy, if there are parameter subspaces that are particularly hard to explore and this is known in advance, it might pay to relatively oversample those parameters so that the decorrelation lengths more closely match those of other parameters. When the hard-to-explore parameters are also fast, there can be substantial overall gains from oversampling them.

Adding oversampling factors for some parameters would alter the optimal parameter order defined in Eq.~\eqref{eq:cost}. If we choose fixed oversampling factors $O_i$ per block $i$, the total cost could be computed in a similar way by simply substituting the vector $\vector{n}$, containing the number of parameters $n_i$ in each block $i$, by a modified vector $\tilde{\vector{n}}$ with components $n_i O_i$ to give the optimal ordering:
\begin{equation}
  \label{eq:costover}
  \hat{\sigma} =
  \operatorname*{arg\,min}_{\sigma\in\mathrm{Sym}(N)} \tilde{\vector{n}}_\sigma^\mathrm{T}\,\tilde{\vector{t}}_\sigma
\end{equation}

The choice of optimal oversampling factors will in general depend on the posterior being sampled, so there is no generally optimal oversampling scheme. In our implementation, we parameterize the oversampling factors as a function of the ratio of evaluation speeds of the different blocks. For simplicity, we refer all ratios to the slowest block $i=1$, use cumulative computation times, and define the oversampling factor as a constant power $o$ of those ratios, rounded to the closest smaller integer:
\begin{equation}
  O_i = \left\lfloor\left(\frac{\tilde{t}_1}{\tilde{t}_i}\right)^{o}\right\rfloor
  \eqp
  \label{overdef}
\end{equation}
With this definition and Eq.\ \eqref{eq:costover}, we can decide on the optimal block ordering for a given global oversampling power $o$.

The oversampling power $o$ is taken as an input parameter to the sampler, and can vary between zero (no oversampling, which will be optimal in the limit in which the fast parameters are independent of the slow parameters and are of no interest in themselves), and larger numbers which progressively increase the number of oversampling steps, investing more computation time per slow step in exchange for better exploration of the fast parameter space. For $o=1$ the same computing time is spent on changing each parameter; even larger values may be useful in cases where the fast parameter space is difficult to explore.

\subsubsection*{Implementation details in \cobaya}

In \cobaya, blocking is automatically determined by the parameter dependencies: input parameters for the same list of components share a common block. Dependencies between likelihood and theory components are automatically calculated based on the requirements for each component, and then finding the components which supply those input requirements (for example, theoretical calculations by one component that are required as input to another). With these dependencies and the input parameters for each component, \cobaya\ then generates the footprint for each parameter block.

Evaluation times per component can be measured at runtime with a small prior sample, or set by hand per theory/likelihood.

Using blocks, footprints and evaluation times, \cobaya minimizes Eq.\ \eqref{eq:cost} by brute force, since in the cases under consideration for us, the number of parameter blocks is $\lesssim 10$.

If oversampling is required, the oversampling power can be specified as an option of the MCMC sampler, and the corresponding optimal ordering is automatically computed. The choice of an optimal oversampling factor depends on the structure of the posterior of the particular problem. We take a default value of $o=0.4$ as a balance between wanting better exploration and not dramatically increasing the computing time. By default we thin the output by the total oversampling factor to avoid outputs becoming very large (and saving output time).

\subsubsection*{Use in dragging}

As an alternative to the simple oversampling scheme defined above, \cobaya also incorporates the possibility of \emph{dragging} along fast directions \cite{Neal04,Lewis:2013hha}. For very large numbers of dragging steps, this sampling method effectively allows slow steps to explore the \emph{marginalized} slow posterior, dragging the fast parameters to the required location of parameter space as the slow parameters are changed. This can be dramatically more efficient where there are strong dependencies between fast and slow parameters (and these are not well captured by the adaptive Cholesky decorrelation). The cost is that it requires two likelihood evaluations per (oversampled) fast step.

Dragging requires a \emph{binary} fast-slow split, whose choice is not trivial when there are more than two blocks. In those cases, \cobaya chooses that split such that it maximizes the log-difference in computation time over all possible fast-slow splits. The given oversampling factor, parameterized as a above in Eq.~\eqref{overdef}, is then applied to the slower block (where $o=1$ now corresponds to spending twice as much time varying each fast parameter as each slow one, corresponding to the two likelihood evaluations required per step). Blocking of parameters within each of the large fast and slow blocks is still used to avoid unnecessary evaluations. Multi-level dragging is not implemented and could be considered in future work, but given the overhead cost is unlikely to be beneficial unless the slow parameters are very slow, since large speed differences would be needed for it to be efficient.

\subsubsection*{Applicability to PolyChord}

\polychord \cite{2015MNRAS.450L..61H,2015MNRAS.453.4384H} is a nested sampling \cite{skilling06} algorithm that utilizes slice sampling \cite{2000physics...9028N} for sampling within iso-likelihood contours. The slice sampling implemented by \polychord allows for decorrelation and parameter speed hierarchy exploitation in a similar way as explained above for MCMC. Parameter blocking and oversampling factors per block can be set automatically using the algorithm described above to sort the blocks and using a global oversampling power parameter to fix oversampling factors (blocking and oversampling can also be set manually).

%%%%%%%%%%%%%%%%%%%%%%%%%%%%%%%%%%%%%%%%%%%%%%%%%%%%%%%%%%%%%%%%%%%%%%%%%%%%%%%%%%%
\section{Source code for the cosmological use case}
\label{sec:cosmo_demo_source}
%%%%%%%%%%%%%%%%%%%%%%%%%%%%%%%%%%%%%%%%%%%%%%%%%%%%%%%%%%%%%%%%%%%%%%%%%%%%%%%%%%%

In this appendix, we reproduce the main parts of the source code necessary to run the use case described in section \ref{sec:cosmo_use_case}. In particular, in figures \ref{fig:src_demo_theory} and \ref{fig:src_demo_likelihood} respectively, we can see the definitions of the Theory and the Likelihood class, and related functions. In figure \ref{fig:src_demo_input_delensed}, we can see the input for the delensed scenario. These and the rest of the necessary files to reproduce the use case (fiducial data generator, covariance matrix and input for the non-delensed case) can be downloaded from \url{https://github.com/CobayaSampler/paper_cosmo_demo}.

\begin{widetext}
\begin{figure*}[ht!]
\begin{python}
import numpy as np
from cobaya.theory import Theory


def feature_envelope(k, c, w):
    """
    Returns the value of the envelope at k. The envelope functional form is

        env(k) = exp(-0.5 (log10(k/c)/w)**2) * sin(2 pi k/l)
    """
    return np.exp(-0.5 * (np.log10(k / c) / w) ** 2)


def feature_power_spectrum(As, ns, A, l, c, w,
                           kmin=1e-6, kmax=10,  # generous, for transfer integrals
                           k_pivot=0.05, n_samples_wavelength=20):
    """
    Creates the primordial scalar power spectrum as a power law plus an oscillatory
    feature of given amplitude A and wavelength l, centred at c with a lognormal envelope
    of log10-width w, as

        Delta P/P = A * exp(-0.5 (log10(k/c)/w)**2) * sin(2 pi k/l)

    The characteristic delta_k is determined by the number of samples per oscillation
    n_samples_wavelength (default: 20).

    Returns a sample of k, P(k)
    """
    # Ensure thin enough sampling at low-k
    delta_k = min(0.0005, l / n_samples_wavelength)
    ks = np.arange(kmin, kmax, delta_k)
    power_law = lambda k: As * (k / k_pivot) ** (ns - 1)
    DeltaP_over_P = lambda k: (A * feature_envelope(k, c, w) * np.sin(2 * np.pi * k / l))
    Pks = power_law(ks) * (1 + DeltaP_over_P(ks))
    return ks, Pks


class FeaturePrimordialPk(Theory):
    """
    Theory class producing a slow-roll-like power spectrum with an enveloped,
    linearly-oscillatory feature on top.
    """

    params = {"As": None, "ns": None,
              "amplitude": None, "wavelength": None, "centre": None, "logwidth": None}

    n_samples_wavelength = 20
    k_pivot = 0.05

    def calculate(self, state, want_derived=True, **params_values_dict):
        As, ns, amplitude, wavelength, centre, logwidth = \
            [params_values_dict[p] for p in
             ["As", "ns", "amplitude", "wavelength", "centre", "logwidth"]]
        ks, Pks = feature_power_spectrum(
            As, ns, amplitude, wavelength, centre, logwidth, kmin=1e-6, kmax=10,
            k_pivot=self.k_pivot, n_samples_wavelength=self.n_samples_wavelength)
        state['primordial_scalar_pk'] = {'k': ks, 'Pk': Pks, 'log_regular': False}

    def get_primordial_scalar_pk(self):
        return self.current_state['primordial_scalar_pk']


def logprior_high_k(A, c, w, k_high=0.25, A_min=5e-3):
    """
    Returns -inf whenever the feature acts at too high k's only, i.e. such that the
    product of amplitude and envelope at `k_high` is smaller than `A_min`, given that the
    envelope is centred at `k > k_high`.
    """
    if c < k_high:
        return 0
    return 0 if A * feature_envelope(k_high, c, w) > A_min else -np.inf

\end{python}
  \caption{
    Source code defining the relevant primordial feature, its \cobaya Theory code wrapper, and the prior described in section \ref{sec:cosmo_use_case}.}
  \label{fig:src_demo_theory}
\end{figure*}
\end{widetext}

%\begin{figure*}[ht!]
%  \centering
%  \lstinputlisting[language=python, basicstyle=\scriptsize, frame=single, linewidth=0.99\linewidth]{src_theory_primordial_Pk.py}
%  \caption{
%    Source code defining the relevant primordial feature, its \cobaya Theory code wrapper, and the prior described in section \ref{sec:cosmo_use_case}.}
%  \label{fig:src_demo_theory}
%\end{figure*}
\begin{widetext}
\begin{figure*}[ht!]
\begin{python}
from cobaya.likelihoods.base_classes import CMBlikes


class SimulatedLikelihood(CMBlikes):
    """
    Likelihood of a partially-delensed CMB survey dataset.
    """

    Alens_delens = None
    dataset_file = None

    def get_requirements(self):
        req = super().get_requirements()
        req['CAMBdata'] = None
        return req

    def logp(self, **data_params):
        cl_dict = cl_dict_from_camb_results(
            self.provider.get_CAMBdata(), self.Alens_delens)
        return self.log_likelihood(cl_dict, **data_params)


def cl_dict_from_camb_results(camb_results, Alens_delens, lmax=None):
    """
    Returns a dictionary of partially-lensed Cl's from a CAMB results object.
    """
    cls = camb_results.get_partially_lensed_cls(Alens_delens, CMB_unit='muK')
    if lmax is None:
        lmax = len(cls[:, 0])
    return {name: cls[:lmax + 1, index]
            for name, index in {"tt": 0, "ee": 1, "bb": 2, "te": 3}.items()}

\end{python}
  \caption{
    Source code defining the likelihood as inheriting from an existing \cobaya base class, \texttt{CMBlikes}. As described in section \ref{sec:cosmo_use_case}, the only methods required are the ones specifying the requirements from the cosmological theory code(s), and the log-likelihood function wrapper that retrieves the (partially) lensed power spectrum and passes it to the routine that performs the actual log-likelihood computation. Notice how the \emph{class attributes} \texttt{Alens\_delens} and \texttt{dataset\_file} can be used in the \texttt{likelihood} block of an input specification (see figure \ref{fig:src_demo_input_delensed}) to create multiple instances of a single Likelihood class, corresponding to the different experiments and delensing scenarios.}
  \label{fig:src_demo_likelihood}
\end{figure*}
\end{widetext}

%\begin{figure*}[ht!]
%  \centering
%  \lstinputlisting[language=python, basicstyle=\scriptsize,  frame=single, linewidth=0.99\linewidth]{src_likelihood_class.py}
%  \caption{
%    Source code defining the likelihood as inheriting from an existing \cobaya base class, \texttt{CMBlikes}. As described in section \ref{sec:cosmo_use_case}, the only methods required are the ones specifying the requirements from the cosmological theory code(s), and the log-likelihood function wrapper that retrieves the (partially) lensed power spectrum and passes it to the routine that performs the actual log-likelihood computation. Notice how the \emph{class attributes} \texttt{Alens\_delens} and \texttt{dataset\_file} can be used in the \texttt{likelihood} block of an input specification (see figure \ref{fig:src_demo_input_delensed}) to create multiple instances of a single Likelihood class, corresponding to the different experiments and delensing scenarios.}
%  \label{fig:src_demo_likelihood}
%\end{figure*}

\begin{figure*}[ht!]
  \centering
\begin{lstlisting}[language=yaml, basicstyle=\scriptsize, frame=single, linewidth=0.99\linewidth, linerange={1-68,105-114}]
theory:
  theory_primordial_Pk.FeaturePrimordialPk:
    python_path: "."
    k_pivot: 0.05
    n_samples_wavelength: 20
  camb:
    external_primordial_pk: True
    extra_args: {
      # Accuracy
      lens_potential_accuracy: 5, Accuracy.IntkAccuracyBoost: 1, Accuracy.lSampleBoost: 3,
      # Below: automatically added by the cosmo-generator
      halofit_version: mead, bbn_predictor: PArthENoPE_880.2_standard.dat,
      num_massive_neutrinos: 1, nnu: 3.046, theta_H0_range: [20, 100]}
likelihood:
  # Change the dataset paths appropriately
  planck_lowl:
    class: likelihood_class.SimulatedLikelihood
    python_path: "."
    dataset_file: /path/to/data/planck_lowl_feature.dataset
    Alens_delens: 0.3
  planck_highl:
    class: likelihood_class.SimulatedLikelihood
    python_path: "."
    dataset_file: /path/to/data/planck_highl_feature.dataset
    Alens_delens: 0.3
  SO_TEB:
    class: likelihood_class.SimulatedLikelihood
    python_path: "."
    dataset_file: /path/to/data/SO_TEB_feature.dataset
    Alens_delens: 0.3
  SO_EB_highl:
    class: likelihood_class.SimulatedLikelihood
    python_path: "."
    dataset_file: /path/to/data/SO_EB_highl_feature.dataset
    Alens_delens: 0.3
params:
  logamplitude:
    prior: [-2, -0.6]
    ref: {dist: norm, loc: -1.1, scale: 0.05}
    proposal: 0.2
    latex: \log_{10}A_\mathrm{feature}
  amplitude:
    value: 'lambda logamplitude: 10**logamplitude'
    latex: A_\mathrm{feature}
  logwavelength:
    prior: [-2.5, -1.8]
    ref: {dist: norm, loc: -2.1, scale: 0.001}
    proposal: 0.0005
    latex: \log_{10}l_\mathrm{feature}
  wavelength:
    value: 'lambda logwavelength: 10**logwavelength'
    latex: l_\mathrm{feature}
  logcentre:
    prior: [-1.15, -0.3]
    ref: {dist: norm, loc: -0.7, scale: 0.008}
    proposal: 0.1
    latex: \log_{10}k_{c,\mathrm{feature}}
  centre:
    value: 'lambda logcentre: 10**logcentre'
    latex: k_{c,\mathrm{feature}}
  logwidth:
    prior: [1e-3, 3]
    ref: {dist: norm, loc: 0.1, scale: 0.02}
    proposal: 0.05
    latex: w_\mathrm{feature}
  # Baseline cosmological parameters
  As:
    value: 'lambda logA: 1e-10*np.exp(logA)'
sampler:
  mcmc:
    covmat: baseline.covmat
    oversample_power: 0.4
    proposal_scale: 1.9
timing: True
output: chains/delensed
\end{lstlisting}
  \caption{
    Cobaya's input in YAML format for the use case described in section \ref{sec:cosmo_use_case} in the delensed scenario. Baseline $\Lambda$CDM cosmological parameters definitions (prior, reference pdf, etc.) are omitted here for brevity.}
  \label{fig:src_demo_input_delensed}
\end{figure*}

%%%%%%%%%%%%%%%%%%%%%%%%%%%%%%%%%%%%%%%%%%%%%%%%%%%%%%%%%%%%%%%%%%%%%%%%%%%%%%%%%%%
%\bibliographystyle{apsrev}
\bibliography{refs,texbase/antony,texbase/cosmomc}
%%%%%%%%%%%%%%%%%%%%%%%%%%%%%%%%%%%%%%%%%%%%%%%%%%%%%%%%%%%%%%%%%%%%%%%%%%%%%%%%%%%

\end{document}